\documentclass{aastex631}

\begin{document}

\title{Galaxy Morphology Classification via Deep Semi-Supervised Learning with Limited Labeled Data}

\correspondingauthor{Zhijian Luo}
\email{zjluo@shnu.edu.cn}

\author[0009-0009-1617-8747]{Zhijian Luo}
\affiliation{Shanghai Key Lab for Astrophysics, Shanghai Normal University, Shanghai 200234, People’s Republic of China}

\author{Jianzhen Chen}
\affiliation{Shanghai Key Lab for Astrophysics, Shanghai Normal University, Shanghai 200234, People’s Republic of China}

\author[0000-0002-2326-0476]{Zhu Chen}
\affiliation{Shanghai Key Lab for Astrophysics, Shanghai Normal University, Shanghai 200234, People’s Republic of China}

\author[0000-0001-8485-2814]{Shaohua Zhang}
\affiliation{Shanghai Key Lab for Astrophysics, Shanghai Normal University, Shanghai 200234, People’s Republic of China}

\author[0000-0003-0688-8445]{Liping Fu}
\affiliation{Shanghai Key Lab for Astrophysics, Shanghai Normal University, Shanghai 200234, People’s Republic of China}
\affiliation{Center for Astronomy and Space Sciences, China Three Gorges University, Yichang 443000, People’s Republic of China}

\author[0000-0001-8244-1229]{Hubing Xiao}
\affiliation{Shanghai Key Lab for Astrophysics, Shanghai Normal University, Shanghai 200234, People’s Republic of China}

\author{Chenggang Shu}
\affiliation{Shanghai Key Lab for Astrophysics, Shanghai Normal University, Shanghai 200234, People’s Republic of China}



\begin{abstract}

Galaxy morphology classification plays a crucial role in understanding the structure and evolution of the universe. With galaxy observation data growing exponentially, machine learning has become a core technology for this classification task. However, traditional machine learning methods predominantly rely on supervised learning frameworks, and their dependence on large of labeled samples limits practical applications. To address this challenge, we propose an innovative hybrid semi-supervised model, GC-SWGAN, designed to tackle galaxy morphology classification under conditions of limited labeled data. This model integrates semi-supervised generative adversarial networks (SGAN) with Wasserstein GAN with gradient penalty (WGAN-GP), establishing a multi-task learning framework. Within this framework, the discriminator and classifier are designed independently while sharing part of the architecture. By collaborating with the generator, the model significantly enhances both classification performance and sample generation capabilities, while also improving convergence and stability during training. Experimental results demonstrate that, on the Galaxy10 DECaLS dataset, GC-SWGAN achieves comparable or even superior classification accuracy (exceeding 75\%) using only one-fifth of the labeled samples typically required by conventional fully supervised methods. Under identical labeled conditions, the model displays excellent generalization performance, attaining approximately 84\% classification accuracy. Notably, in extreme scenarios where only 10\% of the data is labeled, GC-SWGAN still achieves high classification accuracy (over 68\%), fully demonstrating its stability and effectiveness in low-labeled data environments. Furthermore, galaxy images generated by GC-SWGAN are visually similar to real samples. Our approach provides a new solution to the challenge of needing to manually label large amounts of data in astronomy.

\end{abstract}



\section{Introduction} \label{sec:intro}

The morphology of galaxies not only describes their physical appearance but also contains a wealth of important information about their growth processes, including star formation, active galactic nuclei, galaxy mergers, and gas feedback \citep{sandage1986star,hashimoto1998influence,lotz2006rest,parry2009galaxy,buta2011galaxy,rodriguez2017role}. Additionally, galaxy morphology provides crucial insights into the dynamic properties of galaxies, the distribution of stellar populations, and the states of the interstellar medium \citep{holmberg1958photographic,roberts1994physical,deelman2004pegasus,allen2006millennium,benson2010galaxy,conselice2014evolution}. Therefore, understanding galaxy morphology is significant for revealing formation mechanisms, analyzing evolutionary driving factors, and comprehending the evolution of cosmic structures.

Accurate classification of galaxy morphology is a key step toward a comprehensive understanding of galaxy characteristics. With the rapid advancement of observational technologies, such as the Sloan Digital Sky Survey (SDSS) \citep{fukugita1996sloan,york2000sloan}, the Legacy Survey of Space and Time (LSST) \citep{ivezic2019lsst,abell2009lsst}, the Euclid Space Telescope \citep{laureijs2011euclid}, the Kilo-Degree Survey (KiDS) \citep{de2013kilo}, the Dark Energy Survey (DES) \citep{collaboration2016more,abbott2021dark}, the Hubble Space Telescope (HST) \citep{Lallo2012Experience}, and the James Webb Space Telescope (JWST) \citep{2009The}, as well as the upcoming Chinese Space Station Telescope (CSST) \citep{zhan2021wide}, astronomers have accumulated vast amounts of galaxy image data. The scale and complexity of these data far exceed the capabilities of traditional manual classification methods, necessitating the development of automated classification technologies.

Machine learning has emerged as a powerful tool for galaxy morphology classification. Early studies applied classical algorithms, such as Decision Trees \citep{owens1996using}, Support Vector Machines (SVMs) \citep{rouan2008robust,huertas2009robust,aguerri2011revisiting}, Naive Bayes \citep{bazell2001ensembles}, and Locally Weighted Regression \citep{de2004machine}, to relatively small datasets, laying the groundwork for later advancements.

As larger datasets became available, machine learning applications expanded significantly. Researchers employed not only traditional neural networks \citep{storrie1992morphological,lahav1996neural,naim1995automated,goderya2002morphological,ball2004galaxy,de2004machine,banerji2010galaxy} but also more sophisticated Convolutional Neural Networks (CNNs) \citep{dieleman2015rotation,gravet2015catalog,primack2018deep,dominguez2018improving}, demonstrating their effectiveness in handling large-scale data. For example, \citet{cheng2020optimizing} evaluated ten commonly used machine learning methods on the Galaxy Zoo dataset and corrected misclassified labels using supervised learning. The results showed that the CNN model achieved an accuracy of over 99\% in distinguishing elliptical galaxies from spiral galaxies.

Deep learning, with its complex architectures (e.g., multiple convolutional layers, residual blocks), excels at processing high-dimensional data like galaxy images. Studies \citep{dominguez2018improving,zhu2019galaxy,ghosh2020galaxy,walmsley2020galaxy} show that deep learning can efficiently analyze massive datasets, extracting intricate patterns beyond the reach of traditional methods. This not only improves classification accuracy but also reduces human bias, enhancing the objectivity of morphological analysis. For instance, \citet{dieleman2015rotation} developed a deep neural network that leveraged the translational and rotational symmetry of galaxy images, achieving state-of-the-art performance in the Galaxy Challenge competition.

Further algorithmic optimizations and the accumulation of high-quality training data have accelerated progress in galaxy morphology research, providing deeper insights into galactic evolution across cosmic time.

However, existing machine learning methods for galaxy morphology classification predominantly rely on supervised learning, which requires extensive labeled training data. Unfortunately, manual annotation of galaxy morphologies is both time-consuming and costly, hindering the scalability of these methods. For example, although the Galaxy Zoo project has classified hundreds of thousands of galaxies with high-confidence labels through citizen science efforts \citep{willett2013galaxy}, this number is still much smaller compared to the billions of galaxies currently observed.

To address this challenge, the Galaxy Zoo project has continuously expanded its datasets and has accumulated a large amount of labeled data, including Galaxy Zoo Hubble (113,705 galaxies; \citealt{willett2017galaxy}), Galaxy Zoo DECaLS (314,000 galaxies; \citealt{walmsley2022galaxy}), and Galaxy Zoo CANDELS (49,555 galaxies; \citealt{willett2017galaxy}). In addition, there is the Galaxy Zoo DESI sample, which contains 8.7 million galaxies classified using machine learning, with training data provided by volunteers from all previous Galaxy Zoo efforts \citep{walmsley2023galaxy}.

Nevertheless, manual annotation remains a bottleneck for most supervised learning tasks: it not only demands a substantial amount of time and labor but also introduces subjectivity, which can lead to result bias and increased uncertainty in morphological classification. Therefore, developing more efficient automated methods to reduce the dependence on large amounts of labeled data will be an important direction for future research.

Synthetic data represents an alternative method to manual annotation and shows significant potential, particularly in reducing annotation costs. However, synthetic data typically constitute a simplified or simulated version of real data, resulting in discrepancies from the real data distribution. Consequently, any modeling or inference relying on synthetic data carries additional risks. Research indicates that models trained on synthetic data often require fine-tuning with real data prior to deployment to address these distribution discrepancies \citep{jacobs2017finding,jacobs2019finding,pearson2019identifying,ciprijanovic2020deepmerge,he2020deep,krishnakumar2024analysis}.

Generative Adversarial Networks (GANs), introduced by \citet{goodfellow2014generative}, provide a powerful framework for synthetic data generation. A GAN consists of two competing neural networks: a generator that creates synthetic samples from random noise, and a discriminator that distinguishes between real and generated data. Through this adversarial training process, both components progressively improve until reaching equilibrium, the generator produces increasingly realistic outputs while the Discriminator becomes more discerning.

GANs' unsupervised learning capability enables them to automatically extract features and learn data distributions without manual labeling, making them particularly valuable for astronomical applications \citep{schawinski2017generative,ullmo2021encoding,geyer2023deep,das2023generative,gondhalekar2025emulation,luo2024imputation,luo2025cross}. For example, in the field of image enhancement, \citet{schawinski2017generative} used a GAN model to restore morphological features from low-quality galaxy images, while \citet{luo2025cross} utilized a GAN-based Pix2WGAN model to upscale SDSS/DECaLS images to HSC-level resolution; In the area of morphology identification, GAN-generated synthetic galaxies have addressed sample scarcity in morphology classification, improving model performance \citep{krishnakumar2024analysis,yao2024galaxy}.

In recent years, domain adaptation (DA) has emerged as an innovative solution to a core challenge in astronomical deep learning: the degradation of model generalization across datasets caused by variations in observational conditions, instrumental differences, or data distribution shifts. By learning domain-invariant features (i.e., common patterns independent of data sources), DA enables models to maintain high performance on both the source domain (labeled data, such as well-studied sky surveys) and the target domain (unlabeled observations). This capability is particularly valuable for astronomical research, where labeled data are often scarce \citep{vilalta2019general, ciprijanovic2022deepadversaries,ciprijanovic2023deepastrouda, parul2024domain, pandya2025sidda}: DA allows knowledge transfer from existing annotated datasets to new observations, significantly reducing the reliance on costly manual labeling.

Several studies have successfully demonstrated DA's effectiveness in galaxy morphology classification. A notable example is the work of \citet{ciprijanovic2023deepastrouda}, who demonstrated remarkable improvement in classification accuracy by employing a novel domain adaptation framework. They trained their model on one labeled dataset (source domain: SDSS) and one unlabeled dataset (target domain: Galaxy10 DECaLS), achieving up to $40\%$ higher accuracy when analyzing unlabeled target images from the Galaxy10 DECaLS catalog. Similarly, \citet{pandya2025sidda} advanced DA methodologies by developing enhanced algorithms capable of achieving effective domain alignment with minimal hyperparameter tuning and computational overhead. Their approach successfully replicated across multiple simulated and real datasets of varying complexity, establishing their method as a robust solution for practical astrophysical applications.

However, existing methods still have several significant limitations. One approach, GANs can supplement training data by generating synthetic galaxy images; however, these artificial samples often fail to fully capture the complexity of real astronomical observations, diminishing their effectiveness in practical classification tasks. On the other hand, DA techniques, while capable of reducing reliance on annotated data for target domains by transferring knowledge from source domains with well-defined characteristics, are influenced by several factors, including the nature and extent of domain shifts. Additionally, these methods require a sufficient amount of high-quality labeled data from the source domain to effectively train feature mapping and alignment mechanisms.

These limitations are particularly apparent when dealing with rare galaxy types, such as interacting systems, strong gravitational lenses, and ring galaxies. While these objects hold significant scientific value, they often lack sufficient representative samples in major astronomical surveys due to their scarcity and complex observational conditions. This data sparsity severely limits our ability to develop reliable classification models for such intricate phenomena, calling for further methodological innovation and technological advances to address this challenge.

In this study, we propose a novel semi-supervised learning approach that efficiently capitalizes on both limited labeled data and abundant unlabeled observational data, enabling accurate galaxy classification even when labeled data are scarce. Our method, referred to as GC-SWGAN (Semi-Supervised Wasserstein GAN for Galaxy Classification), combines the advantages of semi-supervised generative adversarial networks (SGAN) \citep{odena2016semi} and Wasserstein GAN with Gradient Penalty (WGAN-GP) \citep{adler2018banach}. This hybrid framework offers two key strengths: first, the introduction of the WGAN-GP architecture significantly enhances the stability and convergence of model training; second, it strengthens learning capabilities by leveraging both labeled and unlabeled data, simultaneously maintaining model robustness and further improving classification performance.

To comprehensively evaluate our proposed method, we selected the Galaxy10 DECaLS dataset \citep{leung2019deep} as our experimental platform. This dataset, a carefully curated subset of the Galaxy Zoo project, not only includes strictly labeled samples but also provides a large number of unlabeled galaxy images from DECaLS observations, making it particularly suitable for validating the effectiveness of semi-supervised learning frameworks.

Since the release of the Galaxy10 DECaLS dataset, numerous studies have leveraged it to explore methods for galaxy morphology classification. These studies have employed a variety of advanced techniques from the field of deep learning, including CNNs, residual networks (ResNets), transformers, and transfer learning, achieving significant improvements in galaxy classification accuracy \citep{hui2022galaxy,maile2022equivariance,dagli2023astroformer,pandya20232,yao2024galaxy}. While these efforts have advanced research in galaxy morphology classification, most studies are based on a supervised learning framework, where classification performance heavily depends on the quantity and quality of labeled data.

The Galaxy10 DECaLS dataset comprises approximately 17,700 galaxy samples, which is relatively small for supervised learning tasks. To enhance classification accuracy, previous studies have predominantly allocated over 80\% of the dataset as training data (including test sets and validation sets), reserving less than 20\% for testing purposes. Furthermore, researchers have extensively utilized various data augmentation techniques to expand the training dataset, including reflection, rotation, translation, scaling, shearing, adding Gaussian noise, and Mixup methods \citep{zhang2017mixup,cubuk2020randaugment}. Additionally, some studies have integrated unsupervised generative adversarial networks (GANs) to generate synthetic galaxy images \citep{yao2024galaxy}, thereby increasing the diversity of training samples.

Despite these efforts, the high dependence of supervised learning methods on labeled data significantly limits model performance. For instance, in galaxy classification tasks based on the Galaxy10 DECaLS dataset, the classification accuracy of most models has not exceeded 80\% \citep{maile2022equivariance,huang2024galaxy,yao2024galaxy}. Only a few studies have achieved higher classification accuracy by carefully selecting data augmentation and regularization techniques or adopting complex model architectures, such as deep residual networks combined with transfer learning or hybrid transformer-convolutional architectures \citep{hui2022galaxy,dagli2023astroformer,pandya20232}. The fundamental reason for this phenomenon lies in the scarcity of labeled data, which hampers models’ ability to adequately learn the intrinsic features and distribution patterns of the data.

Our model, GC-SWGAN, offers a more flexible framework through semi-supervised learning. Unlike most previous studies, this method effectively leverages large amounts of unlabeled data for training, significantly reducing the reliance on annotated data. This approach is particularly advantageous in scenarios where annotation costs are high or where labeled data is limited, such as in the search for strong gravitational lenses \citep{stein2022mining}, the detection of galaxy tidal features \citep{desmons2024detecting}, and other similar applications. By employing semi-supervised learning, we not only diminish our dependence on labeled data but also enhance the model’s ability to generalize effectively across unlabeled data.

The organization of this paper is as follows: Section \ref{section:Data} provides an overview of the datasets utilized in this study, including the labeled Galaxy10 DECaLS dataset and the general unlabeled DECaLS images, detailing the characteristics and sources of both. Section \ref{section:Data_pre} presents a comprehensive description of the data preprocessing procedures implemented for model training, which include techniques such as cropping, normalization, data augmentation, and the division into training and testing datasets. Section \ref{section:method} discusses the architecture design and training process of the proposed semi-supervised GC-SWGAN neural network model, emphasizing its innovative features and implementation strategies. Section \ref{sec:results} reports the performance evaluation of the model on the Galaxy10 DECaLS test set, comparing it with traditional supervised learning models to assess its advantages and shortcomings. Finally, Section \ref{section:Summary} summarizes the research findings and discusses potential directions for future improvements in this study.

\section{Data} \label{section:Data}

The labeled galaxy image data we use comes from the Galaxy10 DECaLS dataset, a specialized subset of the Galaxy Zoo project developed by \citet{leung2019deep} for galaxy morphology classification. This dataset encompasses 17,736 labeled color galaxy images across the $g$, $r$, and $z$ bands. All images in this dataset originate from the DESI Legacy Imaging Surveys, which include two DECaLS projects \citep{dey2019overview,walmsley2022galaxy}, the Beijing-Arizona Sky Survey (BASS) \citep{zou2019third}, and the Mayall $z$-band Legacy Survey (MzLS) \citep{dey2019overview}. The galaxy morphology labels are derived from the Galaxy Zoo Data Release 2 \citep{lintott2008galaxy,lintott2011galaxy}. Due to its high-quality annotations and diverse morphological representations, this dataset has been widely utilized in astronomical research.

The Galaxy10 DECaLS dataset provides a comprehensive classification of galaxy morphology, divided into 10 distinct categories. Each class, labeled from 0 to 9, corresponds to specific galaxy types: 0 - Disturbed Galaxies; 1 - Merging Galaxies; 2 - Round Smooth Galaxies; 3 - In-between Round Smooth Galaxies; 4 - Cigar Shaped Smooth Galaxies; 5 - Barred Spiral Galaxies; 6 - Unbarred Tight Spiral Galaxies; 7 - Unbarred Loose Spiral Galaxies; 8 - Edge-on Galaxies without Bulge; and 9 - Edge-on Galaxies with Bulge. This detailed classification enables precise analysis and a better understanding of galaxy morphology.

It is important to note that the Galaxy10 DECaLS dataset exhibits class imbalance, meaning that not all categories have the same number of images. For instance, the “Round Smooth Galaxies” category contains the largest number of labeled images, totaling 2,645, while the “Cigar Shaped Smooth Galaxies” category has only 334 labeled images. This imbalance may affect model training and performance evaluation, necessitating special attention in subsequent research.

\begin{figure} 
        \includegraphics[width=\textwidth]{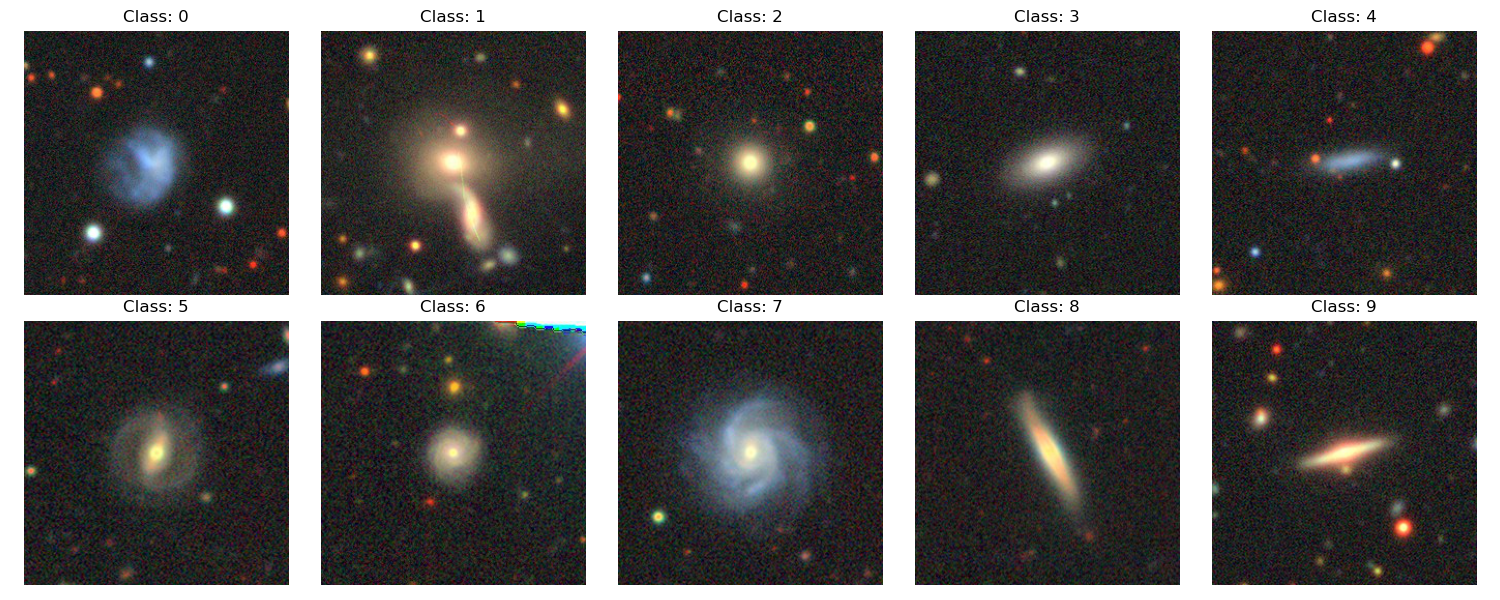} 
        \caption{Random example images of each class in Galaxy10 DECaLS.} 
        \label{fig:G10_DECaLS} 
\end{figure}

Information regarding the specific number of galaxies in each category within the Galaxy10 DECaLS dataset, as well as examples of original galaxy morphology images, can be accessed via the website: https://astronn.readthedocs.io/en/latest/galaxy10.html. In Figure \ref{fig:G10_DECaLS}, we randomly display one original data sample (including galaxy images and labels) from each category in the Galaxy10 DECaLS dataset.

Our unlabeled data also originates from DECaLS \citep{dey2019overview} and shares the same underlying distribution as the Galaxy10 DECaLS images. DECaLS provides approximately two-thirds of the optical imaging coverage for the upcoming Dark Energy Spectroscopic Instrument (DESI), primarily focusing on the Northern Galactic Cap (declination $\leq 32^{\circ}$) and the Southern Galactic Cap (declination $\leq 34^{\circ}$). In the observation process, DECaLS employs a sky-tiling method, conducting three independent observations with slight offsets between each, ranging from approximately $0.1^{\circ}$ to $0.6^{\circ}$. The specific offsets and exposure times are adjusted based on various real-time factors to ensure uniform depth throughout the survey.

DECaLS utilizes the Dark Energy Camera (DECam) \citep{flaugher2015dark}, which is installed on the four-meter Blanco telescope in Chile. The camera features a wide field of view that covers 3.2 square degrees and provides a pixel resolution of 0.262 arcseconds per pixel. The Full Width at Half Maximum (FWHM) values for the $g$, $r$, and $z$ bands are $1.29^{\prime\prime}$, $1.18^{\prime\prime}$, and $1.11^{\prime\prime}$, respectively, resulting in an overall resolution ranging from 0.6 to 1 arcsecond. This high resolution enables DECaLS to resolve finer structures in celestial objects \citep{depoy2008dark}.

Due to limitations in our computational hardware, we randomly selected approximately 18,000 color galaxy images (in the $g$, $r$, and $z$ bands) from the Galaxy Zoo DECaLS catalog \citep{walmsley2022galaxy} as unlabeled samples for model training, matching the number of galaxies in the labeled Galaxy10 DECaLS dataset. 
All galaxy images were downloaded from the Legacy Survey website (https://www.legacysurvey.org/). These selected galaxy images represent a variety of types, ensuring that the model can learn rich morphological features from them. It is noteworthy that the Legacy Survey website contains millions of unlabeled DECaLS galaxy images, providing a vast additional data source for semi-supervised learning. The inclusion of unlabeled data enables our model to learn from a broader range of galaxy features, thus enhancing its generalization capability to unseen data.

\section{Data Preprocessing} \label{section:Data_pre}

To ensure data consistency and improve model performance, we conducted preprocessing on all labeled and unlabeled data. The preprocessing steps include image cropping, pixel value normalization, data augmentation (applied exclusively to labeled data), and dataset splitting. The detailed description of these steps is provided below:

First, we performed image cropping. All downloaded galaxy images are 256×256 pixels, with each main galaxy centered in the image. Considering the limitations of hardware resources, we cropped the image size from 256×256 pixels to 192×192 pixels. This cropping method effectively retains the majority of the main structures of the galaxies while reducing computational resource consumption. The cropped image size remains suitable for model training, ensuring that the model can effectively learn and recognize galaxy features \citep{radford2015unsupervised}.

Second, we conducted pixel value normalization. All pixel values in the downloaded raw images range from [0, 255]. We further normalize these pixel values using the following formula to convert them into the range [-1, 1]:

\begin{equation}
     x^* = \frac{x - 127.5}{127.5}.
     \label{eq:normal} 
\end{equation}
where, $x^{*}$ is the normalized pixel value, $x$ is the original pixel value, and 127.5 is the midpoint of the range from 0 to 255. This normalization method helps accelerate model training and preserves detailed information in the images by reducing the dynamic range while maintaining contrast.

Third, we applied data augmentation. When faced with a low quantity of labeled data, selecting appropriate data augmentation strategies is crucial for enhancing classification model performance. Numerous studies have shown that data augmentation techniques significantly improve classification model performance by increasing the diversity of training data and effectively mitigating overfitting risks. However, in this study, our primary focus is on achieving accurate classification with a limited amount of labeled data by improving model architecture. Therefore, we employed basic geometric transformations, specifically 90-degree rotations, horizontal flips and vertical flips, exclusively on the labeled samples. 
This conservative approach serves two purposes: (1) it provides sufficient data variability to prevent overfitting while (2) maintaining focus on our primary objective of architectural innovation rather than extensive data manipulation. For the unlabeled data drawn from the DECaLS catalog, we deliberately avoided additional augmentation. The catalog's inherent diversity, containing millions of galaxy images with natural variations in orientation and morphology, already provides ample training variation. This decision aligns with our methodological priority for improving the model architecture rather than the complexity of the data processing. 

Finally, we performed dataset splitting. For the unlabeled data, we utilized approximately 18,000 DECaLS galaxy images described in Section \ref{section:Data} for unsupervised model training. Given the abundance of unlabeled images in the DECaLS project, this dataset has significant potential for large-scale expansion. Moreover, images from labeled data can also be treated as unlabeled images and added to the unlabeled dataset for unsupervised training. For the labeled Galaxy10 DECaLS dataset, we implemented several partition schemes to evaluate model performance under varying conditions of labeled data. The first scheme, Train\_test\_91, allocates 90\% of the data for training and 10\% for testing, following common practices in previous studies to assess model performance with relatively sufficient labeled data. The Train\_test\_73 scheme uses 70\% for training and 30\% for testing, further reducing the labeled training data to evaluate performance under medium-high conditions. In the Train\_test\_55 configuration, the dataset is split evenly, with 50\% for training and 50\% for testing, allowing for an assessment of the model in medium labeled data scenarios. Moreover, the Train\_test\_37 scheme uses 30\% for training and 70\% for testing, simulating a low labeled data situation to validate the model’s robustness when labeled data is scarce. The extreme partition, Train\_test\_19, allocates only 10\% of the data for training and 90\% for testing, specifically aimed at exploring the model’s potential under conditions of very minimal labeled data. Through these partition schemes, we aim to comprehensively evaluate the model’s performance in various labeled data contexts.

\section{Methodology} \label{section:method}

We have developed an improved version of the semi-supervised generative adversarial network (SGAN) \citep{odena2016semi} by incorporating Wasserstein GAN with Gradient Penalty (WGAN-GP) \citep{adler2018banach} to enhance loss calculation, thereby improving the model’s stability and convergence. Additionally, we have redesigned the architecture of the discriminator $D$. In traditional SGAN, the discriminator $D$ also serves as the classifier $C$. However, in our improved model, the discriminator and classifier are relatively independent, sharing only partial weights instead of all weights between the discriminator network $D$ and the classifier $C$. This architecture is similar to a dual autoencoder \citep{sutskever2015towards} and allows certain weights to focus on the discrimination task while others specialize in the classification task, thereby enhancing the overall performance of the model. We refer to this improved model as GC-SWGAN. This section will first provide a brief overview of the design principles of SGAN, followed by a detailed description of the architecture of the GC-SWGAN neural network model. Finally, we explain the model’s training process and loss function construction.

\subsection{SGAN Model}

SGAN, proposed by \citet{odena2016semi}, is a deep learning framework that combines Generative Adversarial Networks (GANs) with semi-supervised learning. The core idea is to enhance the model’s overall performance—both classification and generation capabilities—by simultaneously leveraging a small amount of labeled data and a large amount of unlabeled data. In many practical applications, labeled data is often scarce and expensive to obtain, while unlabeled data is relatively abundant and easier to acquire. Consequently, SGAN has been widely applied in scenarios where labeled data is limited.

Similar to traditional GANs, SGAN primarily consists of two components: a generator $G$ and a discriminator $D$. However, unlike traditional GANs, SGAN extends the conventional structure by enabling the discriminator $D$ to not only distinguish between real and fake samples but also classify labeled samples. Specifically, the generator $G$ generates realistic samples from random noise in an attempt to deceive the discriminator $D$. Meanwhile, the discriminator $D$ undertakes two tasks: one is to differentiate between generated samples and real samples (discrimination task), and the other is to classify real samples (classification task).

In traditional GANs, the discriminator $D$ functions solely as a binary classifier, typically outputting an estimated value to indicate whether the input image originates from the real data distribution. This is generally achieved through a feedforward neural network, with the final output being a single sigmoid unit. In contrast, the discriminator $D$ in SGAN acts as a multi-class classifier, employing a softmax output layer that expands the output units to N+1, corresponding to [CLASS-1, CLASS-2, …, CLASS-N, FAKE]. Thus, in this setup, the discriminator $D$ not only performs the discrimination task but also serves as the classifier $C$.

Since the discriminator $D$ can act as the classifier $C$ while executing its discrimination duties, this network model is often referred to as D/C. The advantage of this design is that enhancing the performance of the discriminator $D$ can simultaneously promote the improvement of the classifier $C$, and vice versa. More importantly, this mutually reinforcing relationship also enhances the performance of the generator $G$, thereby improving the quality of the generated samples.

Therefore, SGAN effectively integrates labeled and unlabeled data. By treating unlabeled data as real samples, the training process enhances not only the quality of the generator but also the classification performance of the discriminator, achieving synergistic optimization among $G$, $D$, and $C$. This optimization significantly boosts the model’s performance in both classification and generation tasks, demonstrating its substantial application potential.

\subsection{Hybrid Model GC-SWGAN }

The introduction of SGAN has provided a novel approach to semi-supervised learning. However, the traditional SGAN model, built upon GANs, often exhibits instability during training, which can lead to fluctuations in model performance and even mode collapse, preventing it from fully capturing the comprehensive features of the training data \citep{thanh2020catastrophic}. To address these issues, we have enhanced SGAN by integrating Wasserstein GAN with Gradient Penalty (WGAN-GP) \citep{gulrajani2017improved,chen2021challenges}. By incorporating the Wasserstein distance and adopting a gradient penalty mechanism, we have significantly improved the training stability of semi-supervised learning, effectively enhancing the quality and diversity of generated images while optimizing the performance of both the discriminator and classifier.

In recent years, research has explored the application of this improved SGAN model to various computer vision tasks. For instance, \citet{panwar2019semi} utilized a similar architecture to predict driver states in the field of intelligent driving, while \citet{zeng2023swgan} applied this model in agriculture to assess rice quality. These applications achieved good results. This paper represents the first application of such models in the field of astronomy for galaxy morphology classification, providing a new perspective and methodology for this field.

\begin{figure} 
        \includegraphics[width=\textwidth]{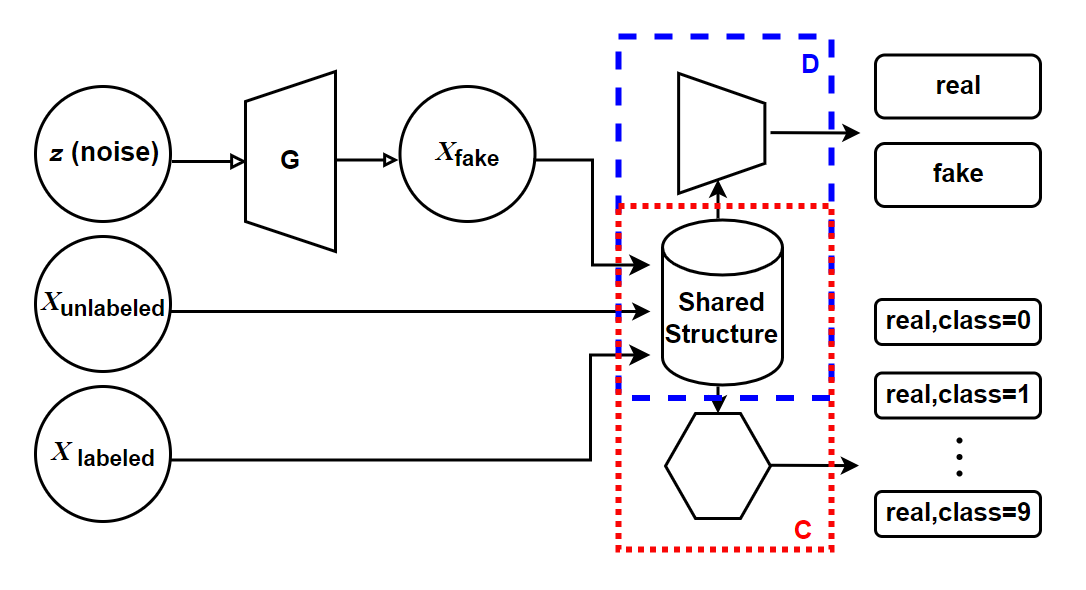} 
        \caption{The network architecture of GC-SWGAN}. The symbol “$z$” denotes the input random noise vector, “$X_{unlabeled}$” represents the input unlabeled real images, “$X_{labeled}$” indicates the input labeled real images, and “$X_{fake}$” refers to the fake images generated by the generator. $G$ is the generator, which employs a multilayer convolutional network. $D$ is the discriminator, whose task is to distinguish between real and fake images. $C$ represents the classifier, which is responsible for learning to assign the correct class labels to real samples. 
        \label{fig:framework} 
\end{figure}

\begin{figure} 
        \includegraphics[width=\textwidth]{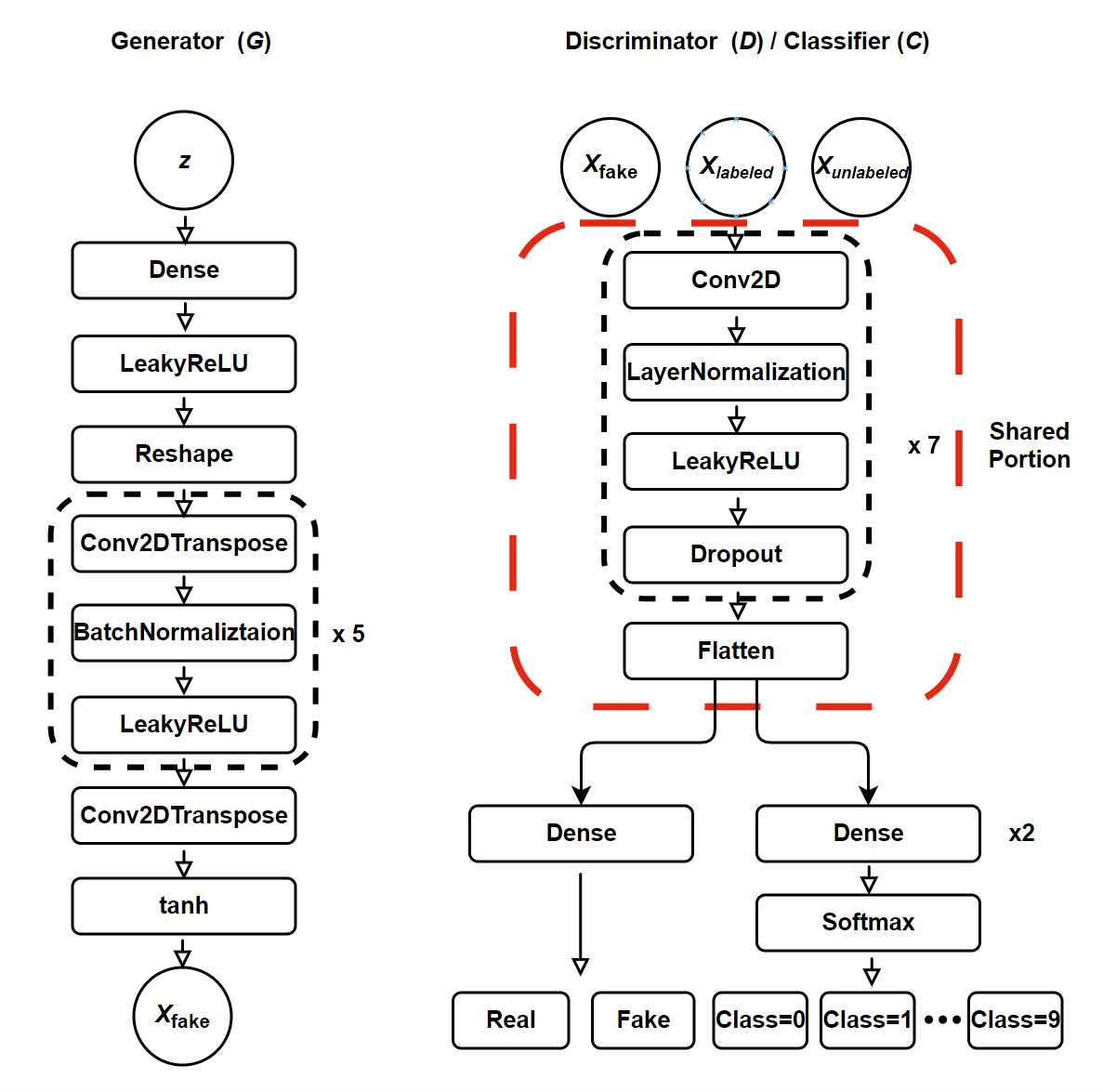} 
        \caption{The detailed structures of the generator $G$, discriminator $D$, and classifier $C$ in the GC-SWGAN} model. 
        \label{fig:G_D_C} 
\end{figure}

Our GC-SWGAN model architecture, as shown in Figure \ref{fig:framework}, consists of three core components: the generator G, discriminator D, and classifier C. Details of their specific structures are illustrated in Figure \ref{fig:G_D_C}. Additionally, we provide comprehensive tabular details describing the architecture of each components (G, D, and C) in the Appendix \ref{section:Appendix} ensuring transparency and reproducibility. Moreover, the code used in this work is publicly available on our GitHub repository \footnote{\url{https://github.com/zjluo-code/GC-SWGAN}} for further exploration and experimentation.

The generator $G$ takes random noise $z$ as input. First, it undergoes initial processing through a fully connected layer, Leaky ReLU activation, and a reshape layer. Subsequently, it is sequentially processed through five identical structural modules. Each module comprises a transpose convolutional layer, batch normalization, and Leaky ReLU activation to progressively reconstruct feature maps. Finally, the generator outputs the generated image through the last transpose convolutional layer with a tanh activation function as the output layer, restricting pixel values within [-1, 1]. This results in an output of a colored image with dimensions 192×192×3.

The discriminator $D$ and classifier $C$ adopt a partially shared network architecture. The shared part consists of seven identical structural modules, each containing a convolutional layer, layer normalization, dropout layer, and Leaky ReLU activation to perform multi-level feature extraction on input data. The non-shared parts use fully connected layers, with an additional softmax layer added for the classifier, enabling them to achieve their respective specific tasks.

This shared mechanism not only reduces model parameters and improves computational efficiency but also allows the discriminator $D$ to focus on distinguishing between real and generated samples, while the classifier $C$ is able to specialize in predicting class labels of real samples.

\subsection{ Model Training and Loss Function}

GC-SWGAN processes data from three different sources:

1) Labeled Real Images ($X_{labeled}$): The discriminator $D$ learns to distinguish between real images and generated images, while the classifier $C$ learns to predict the correct class labels of the labeled real images. The joint training of both components allows the model to simultaneously optimize feature extraction and classification performance. In this study, the Galaxy10 DECaLS dataset serves as the source of labeled real data.

2) Unlabeled Real Images ($X_{unlabeled}$): Only the discriminator $D$ is involved in the training, optimizing its feature extraction capability by assessing the authenticity of the samples.  In this study, given the abundance of unlabeled images in DECaLS, we can only use unlabeled DECaLS images as a source of real unlabeled data. It should be noted that when there is a shortage of unlabeled data, we can also incorporate images from the labeled dataset into the unlabelled dataset to expand the volume of data for unsupervised training.

3) Pseudo Images ($X_{fake}$): The generator $G$ produces pseudo images by inputting random noise vectors $z$. The discriminator $D$ enhances its discriminative ability by identifying these pseudo images as fake data.

While datasets 2) and 3) do not directly contribute to the training of classifier $C$, the shared network components between the classifier and the discriminator mean that optimizing the discriminator will simultaneously improve the classifier. This interplay creates a dynamic feedback loop in the model’s training: improvements in the performance of the discriminator $D$ yield more accurate gradient feedback to the generator $G$, enabling the generation of more representative pseudo-samples. Conversely, enhancements in generator $G$ also boost the feature extraction capabilities of discriminator $D$ through adversarial training, indirectly optimizing the classification performance of classifier $C$.

This collaborative optimization mechanism allows the model to progressively enhance its data generation and classification capabilities, ultimately excelling in the task of galaxy morphology classification.

The loss function of the GC-SWGAN model consists of three components: generator loss $L_G$, discriminator loss $L_D$, and classifier loss $L_C$. The generator loss $L_G$ is an adversarial loss aimed at maximizing the probability that the generated images are recognized as real by the discriminator. In the traditional SGAN, the binary cross-entropy loss (BCE) \footnote{\url{https://www.tensorflow.org/api_docs/python/tf/keras/losses/BinaryCrossentropy}} is used for this calculation. However, in our model, the generator loss is defined using the Wasserstein distance.

The Wasserstein distance \citep{arjovsky2017wasserstein,gulrajani2017improved,adler2018banach}, also known as the Earth Mover’s distance, quantifies the minimum amount of work required to transform one probability distribution into another and serves as a powerful method for measuring the difference between two distributions. A key advantage of this approach is that it offers a smoother and more continuous optimization trajectory, helping to mitigate issues such as vanishing gradients that are often encountered due to JS divergence in certain cases. In the GC-SWGAN model, instead of directly measuring the high-dimensional data distributions, we approximate the Wasserstein distance using the outputs of the discriminator for real and fake images, so the generator loss $L_G$ is defined as:

\begin{equation}
    L_G = -D(G(z)),
\end{equation}
where $G(z)$ is the output of the generator $G$, and $D(G(z))$ is the predicted output of the discriminator $D$ for the generated images.

The loss function $L_D$ consists of two parts. The first part uses the Wasserstein distance to measure the difference between the real samples and the generated samples:
\begin{equation}
    L_{\text{Wasserstein}} = D(x) - D(G(z)),
\end{equation}
where $x$ is the real sample, and $G(z)$ is the generated sample. The second part is the gradient penalty term, which enforces Lipschitz continuity and prevents gradient explosion or vanishing. This term is defined by first creating interpolated samples $x_{\text{interp}}$:
\begin{equation}
    x_{\text{interp}} = \alpha \cdot x + (1 - \alpha) \cdot G(z),
\end{equation}
where $\alpha \in (0, 1)$ is a random number. The gradient penalty is then defined as:
\begin{equation}
    L_{\text{GP}} = \mathbb{E}\left[\left(\left\|\nabla_{x_{\text {interp }}} D\left(x_{\text {interp }}\right)\right\|_2-1\right)^2\right],
\end{equation}
where $\mathbb{E}$ denotes the expectation. The total discriminator loss is finally defined as:
\begin{equation}
    L_D = L_{\text{Wasserstein}} + w_p \cdot L_{\text{GP}} ,
    \label{eq:LD}
\end{equation}
where $w_p$ is the weight for the gradient penalty, which is set to 10 in our study. This choice is based on prior research in the field, particularly the original paper on WGAN-GP \citep{adler2018banach}. The paper points out that a weight of 10 effectively balances training stability and convergence speed. In our experiments, we similarly found that setting the weight to 10 weighs these two factors well, ensuring stability during the training process while avoiding issues such as gradient explosion or vanishing gradients.

For the classifier loss $L_C$,  the output of the classifier employs the softmax function to generate a probability vector, where each element corresponds to the predicted probability of a target class. This loss is calculated as a supervised loss using the categorical cross-entropy function:
\begin{equation}
    L_C=-\sum_{i=1}^Ny_i\log(p_i),
    \label{eq:LC}
\end{equation}
where $y_i$ represents the true class of the input image, $p_i$ is the predicted probability that the input image belongs to class $i$, and $N$ is the total number of categories.

Through the design outlined above, we have constructed a new hybrid semi-supervised galaxy morphology classification model, GC-SWGAN. This model employs an iterative training method, with each iteration encompassing three primary steps: supervised learning, unsupervised learning, and generator training.

During the supervised learning phase, we use the labeled data from the Galaxy10 DECaLS dataset to optimize the classifier. In each iteration, we first randomly select a batch of labeled samples, input them into the classifier, and calculate the cross-entropy loss $L_C$ according to Equation \ref{eq:LC}. Subsequently, we update the parameters of the classifier through backpropagation to enhance its ability to correctly classify galaxy morphologies.

The unsupervised learning phase fully utilizes the abundant unlabeled data to enhance the discriminator's feature extraction capabilities. In this phase, we process two types of samples: (1) real unlabeled images that help the discriminator learn the true data distribution, and (2) generated samples $G(z)$ that improve the discriminator's ability to identify synthetic artifacts. Specifically, for each batch of unlabeled samples, we first randomly generate a corresponding batch of fake samples $G(z)$. Then, according to Equation \ref{eq:LD}, we calculate the discriminator loss $L_D$ that includes the Wasserstein distance and gradient penalty term. By computing and backpropagating the discriminator loss $L_D$, we enable the discriminator to develop more robust decision boundaries without requiring additional labeled data. This phase is particularly crucial for learning generalizable features from the vast amount of available unlabeled astronomical data.

In the generator training phase, we focus on improving the quality of the generator generated sample through adversarial optimization. After generating new samples $G(z')$ from refreshed noise vectors, we evaluate their quality through the discriminator's responses $D(G(z'))$. The generator loss $L_G$ is then minimized to encourage production of more realistic samples. This adversarial push-pull dynamic, constrained by the Wasserstein objective, drives continuous improvement in generation quality while maintaining training stability. 

Through iterative application of these three phases, our model progressively improves both its discriminative and generative capabilities, ultimately leading to better performance on the galaxy classification task. This phased training strategy not only leverages labeled and unlabeled data effectively but also ensures synergistic optimization among the generator, discriminator and classifier. The model demonstrates exceptional stability throughout the training process, successfully evading common pitfalls such as mode collapse and gradient instability typically encountered in traditional GAN training. In the subsequent section, we present the performance of this improved model in galaxy morphology classification tasks. This includes its classification accuracy under low annotation conditions, the calibration performance of the classifier, its application to unlabeled DECaLS data, the quality of generated samples, and comparative analyses with existing models.

\section{Experiments and Results} \label{sec:results}

In this section, we will focus on evaluating the performance of the GC-SWGAN semi-supervised learning method in multi-classification tasks using the Galaxy10 DECaLS dataset, as well as visually assessing the quality of the images generated by the model. The model was implemented using the Keras and TensorFlow 2 libraries \citep{abadi2016tensorflow}, and all experiments were conducted on the NVIDIA L40S GPU platform. During training, the batch size was set to 64. The generator, discriminator, and classifier all utilized the ADAM optimizer \citep{kingma2014adam}, with parameters set to $\beta_1 = 0.5$ and $\beta_2 = 0.999$. The initial learning rate was set to 0.0001 and decayed exponentially after each training iteration with a decay factor of 1/1.000004. The model was trained for a total of 100,000 iterations without implementing any early stopping. This fixed duration was determined empirically to guarantee robust convergence and stability across all experiments. The average training time per experiment was approximately 20 hours.

\subsection{Classification Performance}

We evaluated the classification performance of GC-SWGAN using several commonly used metrics: precision, recall, and F1-score. These metrics provide insights into the model’s performance for each specific category. The corresponding definitions are as follows:
\begin{equation}
    \text{Precision}_i=\frac{\mathrm{TP}_i}{\mathrm{TP}_i+\mathrm{FP}_i},
    \label{eq:precision}
\end{equation}
\begin{equation}
    \text{Recall}_i=\frac{\mathrm{TP}_i}{\mathrm{TP}_i+\mathrm{FN}_i},
\end{equation}
\begin{equation}
    \text{F1-score}_i=2\times\frac{\mathrm{Precision}_i\times\mathrm{Recall}_i}{\mathrm{Precision}_i+\mathrm{Recall}_i},
    \label{eq:F1_score}
\end{equation}
where,$\mathrm{TP}_i$, $\mathrm{FP}_i$, and $\mathrm{FN}_i$ represent the true positives, false positives, and false negatives for category $i$, respectively.

For multi-class classification tasks with balanced classes, it is standard practice to evaluate performance individually for each class and then compute an average. However, due to significant class imbalance in the Galaxy10 DECaLS dataset (as discussed in Section \ref{section:Data}), a weighted averaging approach was employed to ensure fair evaluation across all categories.

The weighted metrics are calculated using:
\begin{equation}
    \text{Weighted Precision}=\frac{\sum_{i=0}^{N-1}(w_i\cdot\mathrm{Precision}_i)}{\sum_{i=0}^{N-1}w_i},
\end{equation}
\begin{equation}
    \text{Weighted Recall}=\frac{\sum_{i=0}^{N-1}(w_i\cdot\mathrm{Recall}_i)}{\sum_{i=0}^{N-1}w_i},
\end{equation}
\begin{equation}
    \text{Weighted F1-score}=\frac{\sum_{i=0}^{N-1}(w_i\cdot\mathrm{F1-score}_i)}{\sum_{i=0}^{N-1}w_i},
\end{equation}
where $w_i$ represents the weight assigned to each category, calculated as:
\begin{equation} w_i = \frac{\text{Number of samples in class } i}{\text{Total number of samples}}. \end{equation}

Additionally, the overall accuracy is defined as:
\begin{equation}
   \mathrm{Accuracy}=\frac{\sum_{i=0}^{N-1}\mathrm{TP}_i}{\sum_{i=0}^{N-1}(\mathrm{TP}_i+\mathrm{FP}_i)},
\end{equation}
which indicates the proportion of correctly classified instances.


We calculated the evaluation metrics on the test sets for different data partitioning schemes, as defined in Section \ref{section:Data}. The results presented in Table \ref{tab:metrics}. Given that the number of samples from the test sets varies between different data partition schemes, and the data distribution and the balance of the classes of the test sets may also change, these factors could potentially affect the comparison of the model performance. To ensure a fair comparison, we further evaluated all models on an identical holdout test set. This holdout test set was defined as the smallest test set from the first experiment (Train\_test\_91), comprising 10\% of the labeled sample size, and was excluded from all training processes in all experiments. The evaluation metrics obtained from this holdout test set are also included in Table \ref{tab:metrics}, highlighted in bold.
The results indicate that the model performance on the holdout test set is consistent with its performance on the test sets of different data partitioning schemes, while showing a slight superiority in all evaluation metrics.

\begin{table}[h]
\centering
\caption{Performance metrics of the model under different data partition schemes. \label{tab:metrics}}
\begin{tabular}{cccccc}
\hline
\hline
Train-Test Split &  Evaluation dataset     & Accuracy  & Weighted Precision  & Weighted Recall & Weighted F1-score   \\ \hline

\multicolumn{6}{c}{\textbf{Single Training for Each Model}}\\ \hline
Train\_test\_91 & Test set  & 83.60\%      & 83.35\%   & 83.60\%  & 83.13\%  \\ 
                & \textbf{Hold-out set}  & \textbf{83.60\%}      & \textbf{83.35\%}   & \textbf{83.60\%}  & \textbf{83.13\%} \\
Train\_test\_73 & Test set  & 81.41\% & 80.96\%  & 81.41\%  & 80.87\%  \\
                & \textbf{Hold-out set} & \textbf{82.58\%} & \textbf{82.17\%} & \textbf{82.58\%} & \textbf{82.08\%}   \\
Train\_test\_55 & Test set     & 80.42\%  & 79.93\%  & 80.42\%  & 79.98\%    \\
                & \textbf{Hold-out set}  & \textbf{81.57\%} & \textbf{81.05\%} & \textbf{81.57\%} & \textbf{81.12\%} \\
Train\_test\_37 & Test set      & 77.21\%  & 77.16\%  & 77.21\%  & 76.63\%  \\ 
                & \textbf{Hold-out set}    & \textbf{78.47\%} & \textbf{78.54\%} & \textbf{78.47\%} & \textbf{78.02\%}   \\ 
Train\_test\_28 & Test set     & 74.68\%  & 74.82\%  & 74.68\% & 74.42\%    \\ 
                & \textbf{Hold-out set}     & \textbf{74.80\%} & \textbf{75.10\%} & \textbf{74.80\%} & \textbf{74.57\%}   \\ 
Train\_test\_19 & Test set    & 68.37\% & 67.85\%  & 68.37\%  & 67.76\%    \\ 
                & \textbf{Hold-out set}  & \textbf{69.00\%} & \textbf{68.54\%} & \textbf{69.00\%} & \textbf{68.31\%}  
                \\\hline
\multicolumn{6}{c}{\textbf{Training Each Model 10 Times }}\\ \hline
Train\_test\_91  &  Test set   & (83.46$\pm$0.37)\%  & (83.30$\pm$0.45)\%  & (83.46$\pm$0.37)\%  & (83.02$\pm$0.42)\%  \\
&\textbf{ Hold-out set} & (\textbf{83.46$\pm$0.37})\% & (\textbf{83.30$\pm$0.45})\%  & (\textbf{83.46$\pm$0.37})\%  & (\textbf{83.02$\pm$0.42})\% \\
Train\_test\_19  & Test set     & (68.74$\pm$0.64)\% & (68.14$\pm$0.83)\% & (68.74$\pm$0.64)\% & (68.00$\pm$0.53)\% \\
    & \textbf{Hold-out set}  & (\textbf{69.00$\pm$0.81})\% & (\textbf{68.50$\pm$0.59})\% & (\textbf{69.00$\pm$0.81})\% & (\textbf{68.28$\pm$0.64})\%
    \\ \hline
\end{tabular}
\end{table}

In addition, to evaluate (or verify) the stability of model training, we conducted further experiments under two extreme data partitioning schemes: one is Train\_test\_91 (where 90\% of the labeled data is used for training), representing the condition of abundant data; the other is Train\_test\_19 (which uses only 10\% of the labeled data for training), simulating a scenario of extreme data scarcity. For each partitioning scheme, we carried out 10 complete training runs, each using a different random seed to ensure the diversity of sample initialization. The lower half of Table \ref{tab:metrics} summarizes the average values and standard deviations of the model’s accuracy, precision, and F1 scores across all experiments. The results indicate that the standard deviations for both partitioning schemes are relatively small, confirming the stability of model training and the consistency of performance under different conditions. Further analysis revealed that the model trained with only 10\% of the labeled data (Train\_test\_19) exhibited a slightly higher standard deviation in performance metrics compared to the model trained with 90\% labeled data (Train\_test\_91). This phenomenon suggests that reducing the proportion of labeled data may have an impact on the stability of model training to some extent.

From the experimental results tabulated in Table \ref{tab:metrics}, it is evident that the semi-supervised model, GC-SWGAN, exhibits strong classification performance across various data partitioning schemes. Notably, under a training-testing ratio where labeled data abundance is relatively high (Train\_test\_91), the model achieved a commendable prediction accuracy of $\sim$ 84\%. This performance is competitive with many state-of-the-art classification methods currently utilized on the Galaxy10 DECaLS dataset.

Previous studies have consistently demonstrated the difficulty of accurately predicting galaxy morphological types using conventional multi-layer CNN networks alone on the Galaxy10 DECaLS dataset. For instance, \citet{huang2024galaxy} reported achieving only approximately 32\% accuracy with such approaches. This limitation has motivated the exploration of more advanced methodologies to enhance predictive performance.

\citet{yao2024galaxy} employed GAN-based synthetic image generation to augment training data samples and evaluated three well-established deep learning architectures - AlexNet \citep{krizhevsky2014one}, VGG \citep{simonyan2014very}, and ResNet \citep{he2016deep} - on the Galaxy10 DECaLS dataset. With a training-to-test set ratio of 9:1, these models demonstrated accuracies ranging from approximately 66\% to 77\%. Additionally, during their image enhancement process, they incorporate partial classification prior information as constraints and perform two types of central cropping based on galaxy type. For edge-on galaxies, the cropping size is 190 × 190 pixels, while for other galaxy types, it is 128 × 128 pixels.

Furthermore, recent studies have also utilized several other commonly used deep learning architectures for supervised learning, such as EfficientNetB0 \citep{tan2019efficientnet}, ConvNext-nano \citep{walmsley2024scaling}, ResNet18 \citep{he2016deep}, and DenseNet121 \citep{huang2017densely}, which achieved top-1 accuracies of 80.9\%, 75.6\%, 73.9\%, and 73.5\% on the Galaxy10 DECaLS morphological classification dataset, respectively \citep{angeloudi2024multimodal}.

While \citet{hui2022galaxy} demonstrated achieving approximately 89\% accuracy using a pre-trained DenseNet-121 model with transfer learning and comprehensive data augmentation techniques in the Galaxy10 DECaLS dataset, their approach heavily relied on transfer learning from prior models. Without this advantage, the model’s performance dropped to around 79\%, still lower than our GC-SWGAN model’s accuracy.

Clearly, transfer learning with pretraining can significantly reduce dependence on large amounts of labeled data and enhance model performance in galaxy classification tasks. For instance, \citet{walmsley2024scaling} utilized the ConvNeXt-nano architecture, pretrained on Galaxy Zoo and ImageNet-12k before fine-tuning on Galaxy10 DECaLS, achieving top-1 accuracies of approximately 89.3\% and 83.9\%, respectively.

Our model differs significantly from prior studies in terms of methodology. While most previous research has depended on supervised learning frameworks, our approach is based on a semi-supervised framework. Consequently, our model does not rely on pre-trained transfer learning but instead achieves good classification performance by leveraging limited labeled data and a large amount of easily accessible unlabeled data.

Additionally, it is worth emphasizing again that \citet{ciprijanovic2023deepastrouda} recently developed a novel domain adaptation technique called DeepAstroUDA. During the training process, this method utilizes only labeled data from the source domain (SDSS) and unlabeled data from the target domain (DECaLS), without using any annotated information from the target domain. Nevertheless, their approach achieved approximately 79\% classification accuracy on the Galaxy10 DECaLS dataset. This result indicates that, like our semi-supervised model, domain adaptation methods are also effective approaches to alleviate the scarcity of labeled data in the target domain.

Though our model’s classification performance is lower than the recent reported best levels (95\%, \citealt{dagli2023astroformer,pandya20232}), its performance remains robust. It must be emphasized that the notable performance of our model can be primarily attributed to the innovative architecture design of the GC-SWGAN framework. During the training phase, we implemented only simple data augmentation techniques (e.g., 90-degree rotations, horizontal and vertical flips) without incorporating sophisticated augmentation strategies or additional regularization mechanisms. This underscores how the success of the model is primarily based on its architectural innovations rather than exhaustive optimization through extensive pre-processing.

As further evidenced by Table \ref{tab:metrics}, GC-SWGAN exhibits consistent and robust performance across a range of annotation conditions: At medium-high annotation ratios (Train\_test\_73), the model achieves approximately 81\% accuracy. Under medium annotation conditions (Train\_test\_55), performance remains strong at around 80\%. Notably, even in more challenging low-annotation scenarios (Train\_test\_37), GC-SWGAN maintains a high level of classification capability with test accuracies exceeding 77\%. Only under extremely limited annotation resources (Train\_test\_19) does performance decline significantly; however, the model still achieving commendable results with accuracy above 68\%.

\begin{figure} 
        \includegraphics[width=\textwidth]{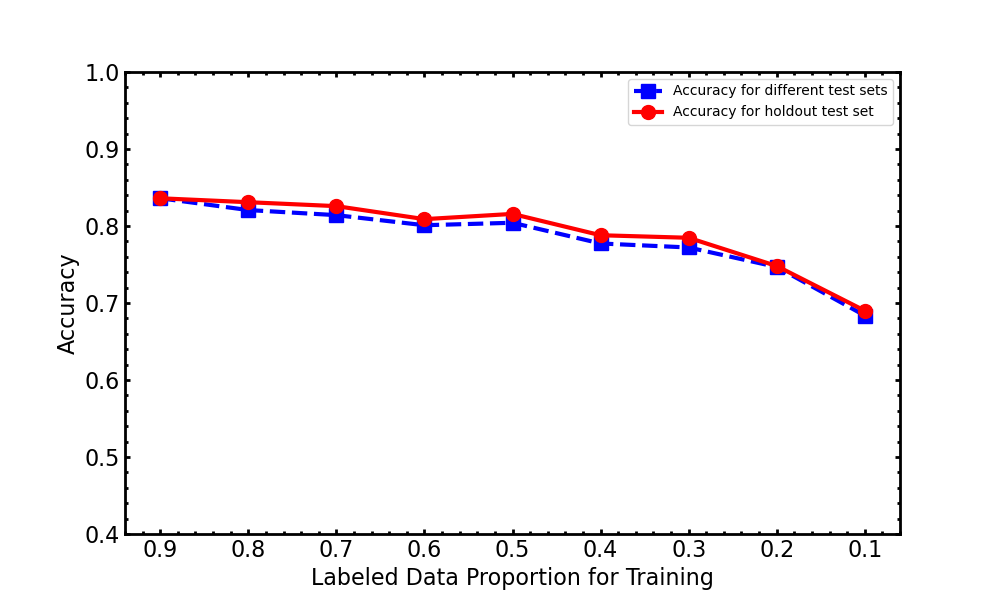} 
        \caption{ Variation in classification accuracy of the GC-SWGAN model with the proportion of labeled data used during training. The red solid line denotes the accuracy on the holdout test set, while the blue dashed line represents the accuracy across the test sets of different data partitioning schemes. } 
        \label{fig:accuracy} 
\end{figure}

Figure \ref{fig:accuracy} illustrates the variation in classification accuracy of the GC-SWGAN model as the proportion of labeled data used during training changes. The red solid line denotes the accuracy on the holdout test set, while the blue dashed line represents the accuracy across the test sets of different data partitioning schemes. From the figure, it can be observed that the trends of the two lines are essentially the same, that is, as the proportion of labeled data decreases, the model’s classification accuracy shows a gradual decline. Specifically, when the proportion of labeled data is above $\sim20\%$, the model’s classification accuracy remains at a high level ($>$75\%), with only minor fluctuations as the labeled data proportion changes. For instance, when the proportion of labeled data decreases from 90\% to 20\%, the classification accuracy only drops slightly from $\sim$84\% to $\sim$75\%, indicating a relatively limited change.

Importantly, it is only when the proportion of labeled data falls below $\sim20\%$ that the model’s classification accuracy begins to decline significantly. This suggests that the GC-SWGAN model is less sensitive to the amount of labeled data, maintaining high classification performance even in scenarios where labeled data is scarce, thus demonstrating good robustness and applicability.

\begin{figure} 
        \includegraphics[width=\textwidth]{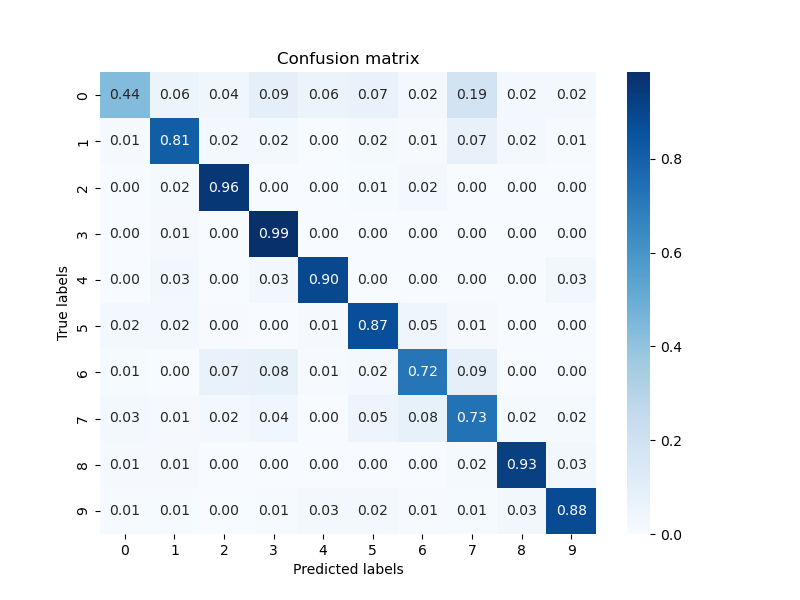} 
        \caption{ Confusion matrix for the target test set under the Train\_test\_91 scheme, illustrating confusion among morphologically similar classes.} 
        \label{fig:cmatrix} 
\end{figure}

To further assess the model’s performance across different categories, Figure \ref{fig:cmatrix} presents the confusion matrix of the final trained model under the Train\_test\_91 condition on the test dataset. The values in the confusion matrix intuitively reflect the proportions of correctly classified and misclassified instances within each category. Additionally, detailed performance metrics for the model across various categories, including precision, recall, and F1-score, are listed in Table \ref{tab:cat_metric}.

Overall, as shown in Figure \ref{fig:cmatrix} and Table \ref{tab:cat_metric}, the GC-SWGAN model demonstrates satisfactory classification performance across most categories under the Train\_test\_91 scheme. However, certain categories with similar visual features still exhibit slight confusion, leading to a decrease in their classification accuracy.

The most significant confusion observed is between “Disturbed Galaxies” and “Unbarred Loose Spiral Galaxies”. It should be noted that while "Disturbed Galaxies" are typically considered to be prone to confusion with "Merging Galaxies" due to their irregular shapes, in our experiments, the confusion between "Disturbed Galaxies" and "Unbarred Loose Spiral Galaxies" stands out more prominently.

This phenomenon may be attributed to the fact that, in the Galaxy10 DECaLS dataset, “Merging Galaxies” often represent systems at earlier stages of merging, typically characterized by having two prominent bright cores, making them easier to distinguish. In contrast, both “Disturbed Galaxies” and “unbarred Loose Spiral Galaxies” exhibit asymmetric morphologies and loose structures, making them more challenging to differentiate.

Moreover, the model faces challenges in distinguishing specific types of spiral galaxies, such as confusion among “Barred Spiral Galaxies,” “Unbarred Tight Spiral Galaxies,” and “Unbarred Loose Spiral Galaxies.” This confusion is understandable since these categories share certain features, thereby increasing the complexity of the classification task.

\begin{table}[h]
\centering
\caption{Performance metrics of the model under the Train\_test\_91 scheme across various galaxy categories. \label{tab:cat_metric}}
\begin{tabular}{ccccc}
\hline
Galaxy Type       & Precision  & Recall & F1-score   \\ \hline
Disturbed Galaxies       & 69.84\% & 43.56\% & 53.66\%  \\
Merging Galaxies       & 83.44\% & 80.86\% & 82.13\% \\
Round Smooth Galaxies       & 89.86\% & 95.75\% & 92.71\%   \\
In\_between Round Smooth Galaxies       & 82.85\% & 98.51\% & 90.00\%    \\ 
Cigar Shaped Smooth Galaxies       & 61.90\% & 89.66\% & 73.24\%  \\ 
Barred Spiral Galaxies       & 83.89\% & 87.19\% & 85.51\%  \\
Unbarred Tight Spiral Galaxies       & 77.14\% & 72.19\% & 74.59\%  \\
Unbarred Loose Spiral Galaxies       & 78.04\% & 73.16\% & 75.52\%   \\
Edge-on Galaxies without Bulge       & 89.88\% & 92.64\% & 91.24\%    \\ 
Edge-on Galaxies with Bulge      & 92.55\% & 88.32\% & 90.39\%   \\ \hline

\end{tabular}
\end{table}

\subsection{ Calibration Performance of Classifier}

The classifier of the GC-SWGAN model uses the Softmax activation function as the output layer, converting the raw outputs (logits) of the neural network into a probability distribution across categories, ensuring that the sum of these probability values equals 1. Therefore, the model can not only predict the categories of the input galaxy images but also output the predicted probability values for each category. Although our experiments have shown that the probabilities based on these Softmax functions can effectively contribute to decision-making in the multi-class classification task of galaxy morphology (with the category having the highest probability being selected as the final prediction result in this study), these probability values may deviate from actual situations. Specifically, predicted probabilities can be influenced by factors such as the distribution of training data, network architecture, and the training process, leading to discrepancies between predicted results and true probabilities.

To evaluate the probability calibration of the classifier in the GC-SWGAN model, we selected two commonly used calibration metrics: Expected Calibration Error (ECE) and Brier Score. 
ECE is a crucial metric for assessing how well the predicted confidence from a classification model aligns with the actual real-world probabilities. Specifically, ECE measures the degree of calibration between the predicted probabilities of the model and the observed probabilities \citep{naeini2015obtaining,guo2017calibration}. The computation involves dividing the predicted probability range into intervals (bins) and calculating the weighted absolute difference between the average predicted probability within each bin and the true positive rate for that bin. This process ultimately yields an ECE value ranged from [0,1], with lower values indicating better calibration performance.

The Brier Score is another important metric for assessing the accuracy of probability predictions, measuring the mean squared difference between predictive probabilities and actual outcomes \citep{brier1950verification,minderer2021revisiting}. Specifically, for each sample, the square of the difference between its predicted probabilities across various categories and the corresponding actual category (represented as 0 or 1 to indicate whether it belongs to the positive class) is computed, and the average of these squared differences across all samples yields the Brier Score. Similar to ECE, the Brier Score also ranges from [0,1], with lower values indicating more accurate predictions by the model.

\begin{figure} 
        \includegraphics[width=\textwidth]{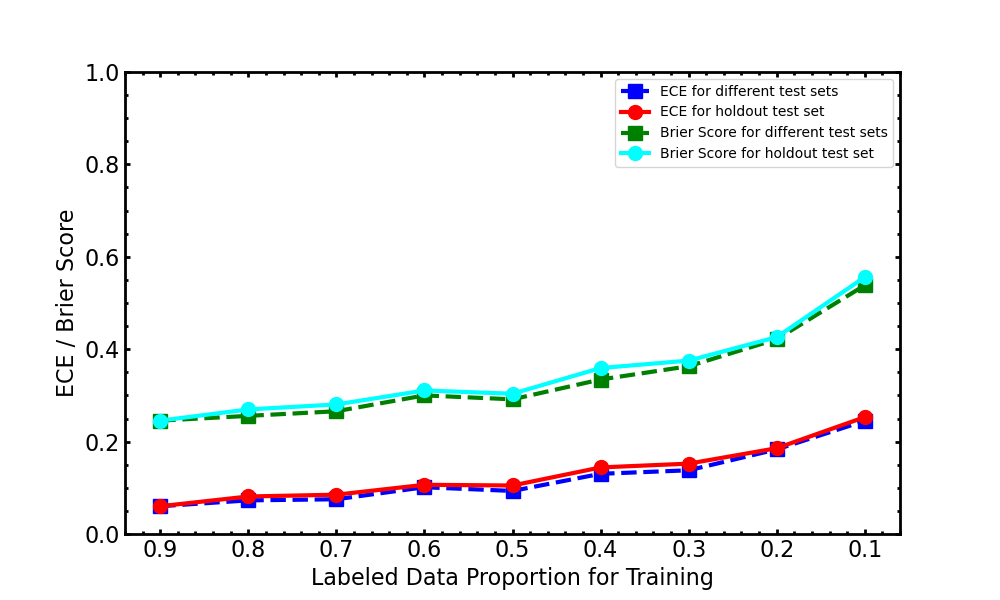} 
        \caption{ Variation of ECE and Brier score with the proportion of labeled data used during training. The solid lines (red for ECE, cyan for Brier score) correspond to results obtained on the holdout test set, while  the dashed lines (blue for ECE, green for Brier score) show variations in these metrics across the test sets of different data partitioning schemes.} 
        \label{fig:ece_brier} 
\end{figure}

We calculated the ECE and Brier score for models trained under different training-testing set partitioning schemes on their respective test sets and the holdout dataset. Figure \ref{fig:ece_brier} illustrates the variations of ECE and Brier score with the proportion of labeled data used during model training. The solid red line and dashed blue line represent the ECE values, while the solid cyan line and dashed green line represent the Brier scores. Specifically, the solid red line (ECE) and solid cyan line (Brier score) correspond to the results obtained on the holdout test set, while the dashed blue line (ECE) and dashed green line (Brier score) show the changes in these metrics across different testing datasets under varying data partitioning schemes. These lines clearly demonstrate the impact of differences in the proportion of labeled data used for training on the calibration performance of the model.

As an example, based on the evaluation of a holdout test set: Under the Train\_test\_91 split scheme (where labeled data is relatively abundant), the model demonstrates good calibration with an ECE of 0.065 and a Brier score of 0.245, indicating that the predicted probabilities are reasonably accurate and reliable. However, as the proportion of labeled data gradually decreases, both the ECE and Brier score show an upward trend. When the labeled data ratio drops to 10\% (Train\_test\_19), these metrics increase to 0.245 and 0.538, respectively, suggesting that the model’s calibration performance and probability prediction accuracy are significantly impacted when labeled data is extremely scarce. Future research could explore methods for optimizing the calibration of model outputs to enhance their reliability and applicability in real-world scenarios.

\subsection{Application of Classifiers on Unlabeled DECaLS Data}

In this subsection, we attempt to directly apply the well-trained GC-SWGAN classifier to unlabeled DECaLS data for class prediction. Due to the lack of class labels, a quantitative evaluation is not feasible. Therefore, we only perform visual validation on the classifications of some unlabeled DECaLS images to assess the applicability of the model on this dataset.

It should be noted that the Galaxy10 DECaLS dataset is a subset of DECaLS and contains samples where galaxy morphology can be unambiguously identified by volunteers. As stated in the Galaxy10 DECaLS documentation \footnote{https://astronn.readthedocs.io/en/lateest/galaxy10.html}, “Galaxy10 only contains images for which more than 55\% of the votes agree on the class”. This means that for the 10 categories, an image is considered to belong to a specific category only if it receives over 55\% agreement from volunteers. If no category meets this consensus threshold (over 55\%), the image will not be included in Galaxy10 DECaLS due to the lack of clear consensus.

Compared to Galaxy10 DECaLS, unlabeled DECaLS images exhibit greater diversity and complexity in galaxy morphology. Directly applying a model trained on Galaxy10 DECaLS to the unlabeled DECaLS dataset may limit the model’s classification ability, resulting in less accurate or comprehensive classification outcomes. Therefore, the application of GC-SWGAN on unlabeled DECaLS images is limited to providing preliminary classifications for this dataset. To ensure classification accuracy, human review of the model’s outputs is still necessary, along with domain expertise to validate classification accuracy and address potential omissions or misclassifications by the model.

\begin{figure} 
        \includegraphics[width=\textwidth]{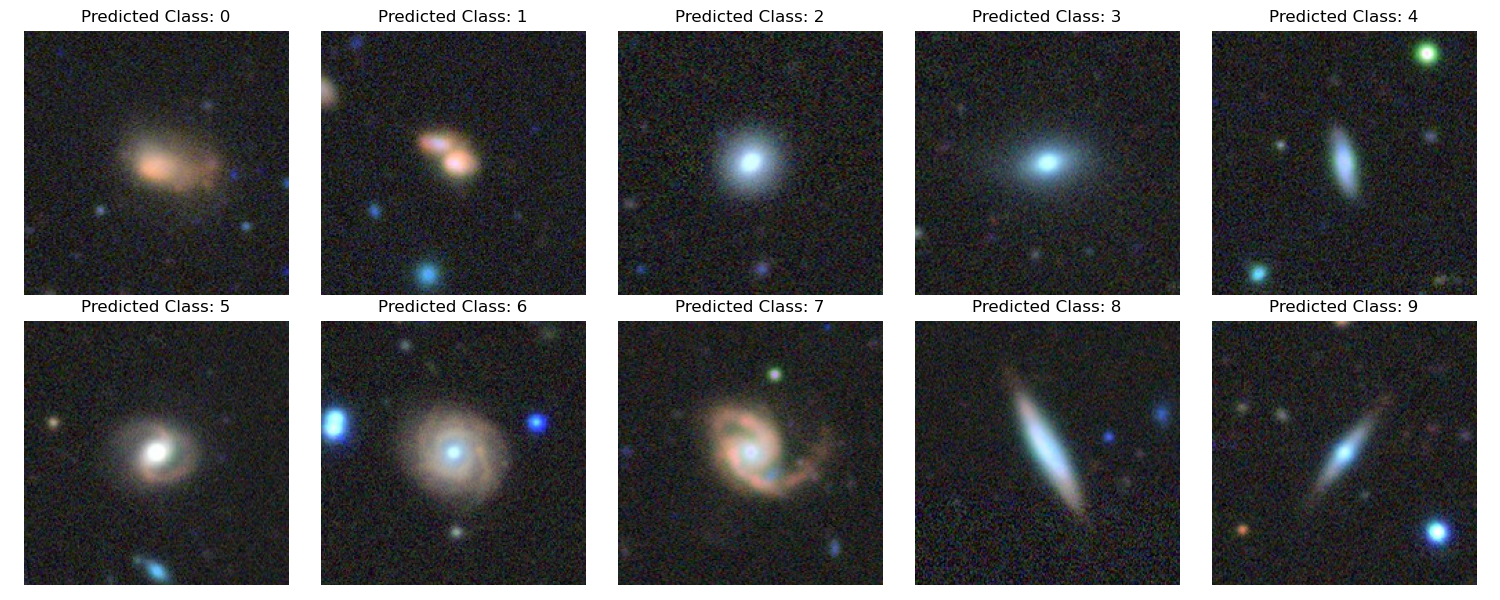} 
        \caption{ Representative images of predicted categories in unlabeled DECaLS dataset.} 
        \label{fig:unlabeled_examples} 
\end{figure}

Figure \ref{fig:unlabeled_examples} presents examples of unlabeled DECaLS galaxy images classified by the GC-SWGAN classifier, displaying one representative image for each category. Through a visual inspection of the classifications of several hundred unlabeled DECaLS galaxy images, we found that the model is capable of providing preliminary morphological classifications for new data, further confirming its strong generalization capability. Future research could consider incorporating more diverse training data and fine-tuning the model to further improve classification accuracy and robustness.

\begin{figure} 
        \includegraphics[width=\textwidth]{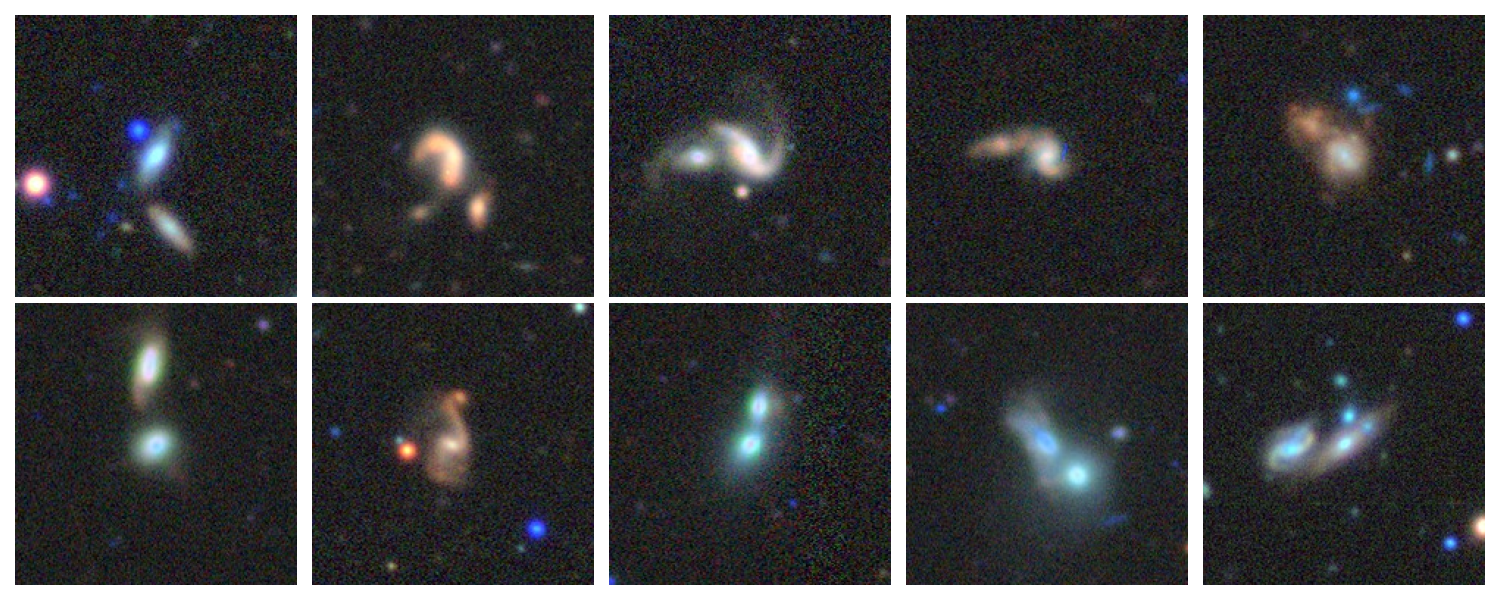} 
        \caption{ Some example images of Merging Galaxies in unlabeled DECaLS dataset.} 
        \label{fig:merging} 
\end{figure}

Additionally, Table \ref{tab:cat_metric} reveals that the model excels in most predefined categories. Taking “Round Smooth Galaxies" as an example, the model achieves classification precision, recall, and F1-scores of 89.86\%, 95.75\%, and 92.71\%, respectively, for this category. For “Merging Galaxies", the model's classification precision, recall, and F1-scores also reach 83.44\%, 80.86\%, and 82.13\%, respectively. These results indicate that GC-SWGAN can be effectively applied to unlabeled DECaLS samples to assist in screening specific types of galaxies. Figure \ref{fig:merging} presents examples of “Merging Galaxies" selected by this classifier, showcasing diverse morphological features such as irregular bridges, tidal tails, and varying degrees of asymmetry. Future research could further analyze these selected galaxy samples to explore the underlying physical mechanisms and evolutionary histories.

\subsection{Visual Analysis of Generated Images}

In our GC-SWGAN model, a collaborative optimization mechanism exists among the generator $G$, discriminator $D$, and classifier $C$. Through this mechanism, the model achieves significant improvements in both classification performance and image generation quality. Specifically, the generator learns more comprehensive feature representations from feedback provided by the discriminator and classifier, enabling it to produce higher-quality samples.

\begin{figure} 
        \includegraphics[width=\textwidth]{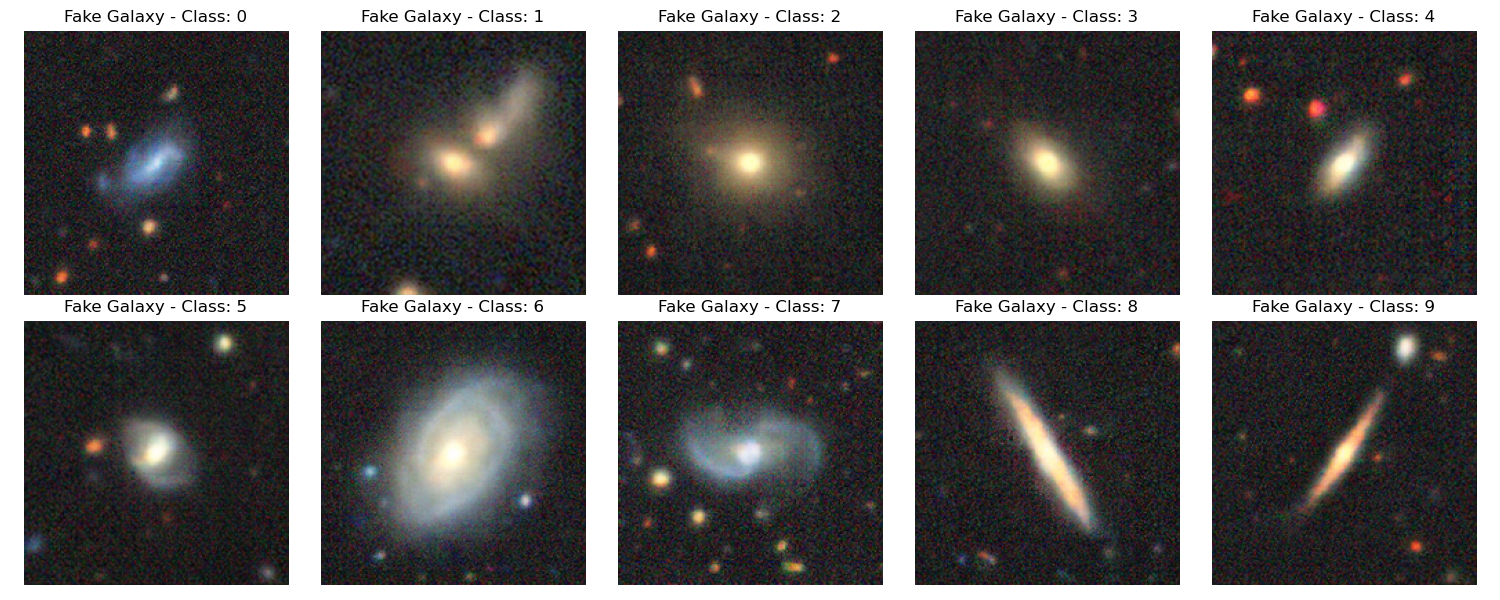} 
        \caption{ Randomly selected samples from each category in the generated fake galaxies produced by the generator.} 
        \label{fig:predfig} 
\end{figure}

While evaluating the quality of generated images is not the primary focus of this study, we conducted a simple visual inspection to assess their realism. Figure \ref{fig:predfig} shows randomly selected samples from each category produced by the generator, including fake galaxy images with predicted labels. These examples demonstrate that images generated using the GC-SWGAN method are visually highly similar to real galaxies.

The generated samples exhibit remarkable diversity in morphological types, such as spiral galaxies, elliptical galaxies, and irregular galaxies. In terms of detailed features, the generated galaxy images closely resemble real galaxies in terms of morphology, brightness, and structure. Notable features like spiral arms, galactic nuclei, and dust lanes are effectively captured. Based on this visual analysis, the generated samples are nearly indistinguishable from real galaxies, further validating the generator’s high-quality generation capability. This advanced generation ability also enhances the discriminator’s feature extraction capabilities through adversarial training, thereby optimizing the classifier’s classification performance.

Furthermore, by expanding the dataset for model training with these high-quality images generated by the generator, we can enhance the model’s ability to identify galaxy morphologies with extremely high diversity and improve its robustness. This method exhibits promising potential, and we believe it is a direction worth further in-depth research and exploration.

\section{Summary} \label{section:Summary}

We propose a new hybrid semi-supervised model, GC-SWGAN, to address the challenge of galaxy morphology classification under conditions of limited labeled data. This model is trained by combining a small labeled dataset with a large collection of unlabeled data, resulting in significant improvements in classification performance. Our model is based on the traditional SGAN architecture and is optimized by incorporating Wasserstein GAN along with its improved version, WGAN-GP (Wasserstein GAN with Gradient Penalty).

By introducing Wasserstein distance and gradient penalty mechanisms, we significantly enhance the training stability of the semi-supervised learning process while improving the quality and diversity of generated images. Additionally, we optimize the designs of the discriminator and classifier by employing a method that allows for their independent architecture while sharing some components of the network. Through the collaborative optimization of the discriminator, generator, and classifier, the model’s classification performance and generative capability have been significantly enhanced.

The results of our experiments on the Galaxy10 DECaLS dataset demonstrate that GC-SWGAN excels in the galaxy classification task. When sufficient labeled data is available (e.g., using 90\% of the data for training), the overall classification accuracy reaches approximately 84\%, surpassing most existing supervised learning methods. Furthermore, even with limited labeled data, GC-SWGAN maintains strong performance, achieving results comparable to many supervised methods with only one-fifth of the labeled data. In situations with very scarce labeled data (e.g., using only 10\% of the labeled data), the model still achieves high classification accuracy. Analysis of the confusion matrix and detailed performance metrics across various categories further indicates that the model performs well in most categories, with confusion occurring primarily between visually similar categories. This demonstrates its effectiveness in feature extraction and category differentiation.

The well-trained GC-SWGAN classifier can be directly applied to the unlabeled DECaLS dataset for preliminary type predictions. Our results show that the model performs well on new data and can effectively assist in identifying specific types of galaxies, fully demonstrating its strong generalization capability. Future research could further analyze the selected samples to explore the underlying physical mechanisms and evolutionary processes, thereby enhancing our understanding of galaxy morphology and evolution.

In addition, the GC-SWGAN model also exhibits high quality in generating galaxy images, with the generated galaxy samples showing high consistency with real galaxies in terms of morphology, brightness, and structure, as well as demonstrating rich diversity. This result indicates that the GC-SWGAN, after multiple iterations of training, can generate high-quality synthetic samples.

However, despite the good performance of the GC-SWGAN model, there is still potential for further enhancement. Due to computational resource limitations, the amount of unlabeled data used during training is currently equal to that of the labeled data. Future research could explore incorporating larger quantities of unlabeled image data to facilitate better learning of the intrinsic structures and distribution characteristics of the images, thus improving classification accuracy and generalization ability. Additionally, employing more data augmentation techniques, such as Mixup \citep{zhang2017mixup} and RandAugment \citep{cubuk2020randaugment}, could help expand the labeled dataset and enhance the model’s robustness and effectiveness. Furthermore, implementing advanced input image preprocessing techniques, such as morphological opening operations \citep{hui2022galaxy} and increasing the resolution of input images \citep{luo2025cross}, could enable the model to extract features more effectively and fully leverage its potential.

\section{Acknowledgments}

Z.J.L. acknowledges the support from the Shanghai Science and Technology Foundation Fund (Grant No. 20070502400) and the scientific research grants from the China Manned Space Project with Grand No. CMS-CSST-2025-A07. S.H.Z. acknowledges support from the National Natural Science Foundation of China (Grant No. 12173026), the National Key Research and Development Program of China (Grant No. 2022YFC2807303), the Shanghai Science and Technology Fund (Grant No. 23010503900), the Program for Professor of Special Appointment (Eastern Scholar) at Shanghai Institutions of Higher Learning, and the Shuguang Program (23SG39) of the Shanghai Education Development Foundation and Shanghai Municipal Education Commission. L.P.F. acknowledges the support from the National Natural Science Foundation of China (NSFC 11933002). H.B.X. acknowledges the support from the National Natural Science Foundation of China (NSFC 12203034) and the Shanghai Science and Technology Fund (22YF1431500). This work is also supported by the National Natural Science Foundation of China under Grant No. 12141302.

The DESI Legacy Imaging Surveys consist of three individual projects: the Dark Energy Camera Legacy Survey (DECaLS), the Beijing-Arizona Sky Survey (BASS), and the Mayall z-band Legacy Survey (MzLS). These surveys utilized facilities such as the Blanco, Bok, and Mayall telescopes, supported by the National Science Foundation (NSF) and operated by different observatories including NSF’s NOIRLab.

We thank the respective teams and funding agencies for making these data publicly available. For detailed acknowledgments and funding information, please refer to the original publications and data release notes. 

We also extend our sincere appreciation to the anonymous referee for their valuable comments and constructive suggestions, which have substantially enhanced the rigor and clarity of this manuscript.


%




{\color{red}
\appendix
\section{Model Architectures} \label{section:Appendix}

\subsection{The Architecture of the Generator G} 

Table \ref{table:generator_g} presents the detailed architecture of the generator G, including the names of each layer, the number of parameters, the corresponding output shapes, and other properties such as filter size, kernel size and stride.

\begin{table}[h]
\centering
\caption{ Architecture of Generator G in the GC-SWGAN Model}
\label{table:generator_g}
\begin{tabular}{l c c l}
\hline
\hline
\textbf{Layer} & \textbf{Output Shape} & \textbf{Parameters} & \textbf{properties} \\ \hline
Input &  & 0 & Shape: (100,) \\ 
Dense & (9216,) & 921600 & Filters:9216 \\ 
LeakyReLU & (9216,) & 0 & Activation:LeakyReLU, Alpha:0.2 \\ 
Reshape & (6, 6, 256) & 0 & Reshaping for convolutional input \\ 
Conv2DTranspose & (6, 6, 256) & 1048576 & Filter:256, Kernel:4x4, Stride:1 \\ 
BatchNormalization & (6, 6, 256) & 1024 & Normalization layer \\ 
LeakyReLU & (6, 6, 256) & 0 & Activation:LeakyReLU, Alpha:0.2 \\ 
Conv2DTranspose & (12, 12, 256) & 1048576 & Filter:256, Kernel:4x4, Stride:2 \\ 
BatchNormalization & (12, 12, 256) & 1024 & Normalization layer \\ 
LeakyReLU & (12, 12, 256) & 0 & Activation:LeakyReLU, Alpha:0.2 \\ 
Conv2DTranspose & (24, 24, 128) & 524288 & Filter:128, Kernel:4x4, Stride:2 \\ 
BatchNormalization & (24, 24, 128) & 512 & Normalization layer \\ 
LeakyReLU & (24, 24, 128) & 0 & Activation:LeakyReLU, Alpha:0.2 \\ 
Conv2DTranspose & (48, 48, 128) & 262144 & Filter:128, Kernel:4x4, Stride:2 \\ 
BatchNormalization & (48, 48, 128) & 512 & Normalization layer \\ 
LeakyReLU & (48, 48, 128) & 0 & Activation:LeakyReLU, Alpha:0.2 \\ 
Conv2DTranspose & (96, 96, 64) & 131072 & Filter:64, Kernel:4x4, Stride:2 \\ 
BatchNormalization & (96, 96, 64) & 256 & Normalization layer \\ 
LeakyReLU & (96, 96, 64) & 0 & Activation:LeakyReLU, Alpha:0.2 \\ 
Conv2DTranspose & (192, 192, 3) & 3072 & Filter:1, Kernel:4x4, Stride:2 \\
Tanh & (192, 192, 3) & 0 & Activation: tanh \\
\hline
\end{tabular}
\end{table}

\subsection{The Shared Architecture Between Discriminator D and Classifier C}

Table \ref{table:model_shared} presents the detailed architecture of the shared portion between generator G and classifier C, including the names of each layer, the number of parameters, the corresponding output shapes, and other properties such as filter size, kernel size and stride.

\begin{table}[h]
\centering
\caption{The shared architecture between Discriminator D and Classifier C}
\label{table:model_shared}
\begin{tabular}{l c c l}
\hline
\hline
\textbf{Layer} & \textbf{Output Shape} & \textbf{Parameters} & \textbf{Properties} \\ 
\hline
Input &  & 0 & (192,192,3) \\ 
Conv2D & (192, 192, 64) & 3072 & Filter:64, Kernel:4x4, Stride:1  \\ 
LayerNormalization & (192, 192, 64) & 128 & Normalization layer\\ 
LeakyReLU & (192, 192, 64) & 0 & Activation: LeakyReLU, Alpha:0.2 \\ 
Dropout & (192, 192, 64) & 0 & Regularization, Rate:0.15 \\ 
Conv2D & (96, 96, 64) & 65536 & Filter:64, Kernel:4x4, Stride:2 \\ 
LayerNormalization & (96, 96, 64) & 128 & Normalization layer \\ 
LeakyReLU & (96, 96, 64) & 0 & Activation: LeakyReLU, Alpha:0.2 \\ 
Dropout & (96, 96, 64) & 0 & Regularization, Rate:0.15 \\ 
Conv2D & (48, 48, 128) & 131072 & Filter:128, Kernel:4x4, Stride:2 \\ 
LayerNormalization & (48, 48, 128) & 256 & Normalization layer \\ 
LeakyReLU & (48, 48, 128) & 0 & Activation: LeakyReLU, Alpha:0.2 \\ 
Dropout & (48, 48, 128) & 0 & Regularization, Rate:0.15 \\ 
Conv2D & (24, 24, 128) & 262144 & Filter:128, Kernel:4x4, Stride:2 \\ 
LayerNormalization & (24, 24, 128) & 256 & Normalization layer\\ 
LeakyReLU & (24, 24, 128) & 0 & Activation: LeakyReLU, Alpha:0.2 \\ 
Dropout & (24, 24, 128) & 0 & Regularization, Rate:0.15 \\ 
Conv2D & (12, 12, 256) & 524288 & Filter:256, Kernel:4x4, Stride:2 \\ 
LayerNormalization & (12, 12, 256) & 512 & Normalization layer \\ 
LeakyReLU & (12, 12, 256) & 0 & Activation: LeakyReLU, Alpha:0.2 \\ 
Dropout & (12, 12, 256) & 0 & Regularization, Rate:0.15 \\ 
Conv2D)& (6, 6, 256) & 1048576 & Filter:256, Kernel:4x4, Stride:2 \\ 
LayerNormalization & (6, 6, 256) & 512 & Normalization layer\\ 
LeakyReLU & (6, 6, 256) & 0 & Activation: LeakyReLU, Alpha:0.2 \\ 
Dropout & (6, 6, 256) & 0 & Regularization, Rate:0.15 \\ 
Conv2D & (6, 6, 256) & 1048576 & Filter:256, Kernel:4x4, Stride:2 \\ 
LayerNormalization & (6, 6, 256) & 512 & Normalization layer \\ 
LeakyReLU & (6, 6, 256) & 0 & Activation: LeakyReLU, Alpha:0.2 \\ 
Flatten & (9216,) & 0 & Flattening layer \\ 
Dropout & (9216,) & 0 & Regularization, rate:0.15 \\ \hline
\end{tabular}
\end{table}

\subsection{The Unique Non-shared Architectures of Discriminator D and Classifier C}

Table \ref{table:nonshared} presents the detailed architectures of the unique non-shared portions of discriminator D and classifier C, including the names of each layer, the number of parameters, the corresponding output shapes, and other properties such as filter size, kernel size and stride.

\begin{table}[h]
\centering
\caption{The unique Non-shared architectures of discriminator D and classifier C}
\label{table:nonshared}
\begin{tabular}{lcccl}
\hline
\hline
\textbf{Model} & \textbf{Layer} & \textbf{Output Shape} & \textbf{Parameters} & \textbf{Properties} \\ \hline
Discriminator & Input &  & 0 & (9216,) \\  
& Dense & (1,) & 9217 & Filter:1 \\  \hline
Classifier & Input &  & 0 & (9126,) \\ 
 & Dense & (128,) & 1179116 & Filter:128 \\
 & LeakyReLU & (128,) & 0 & Activation: LeakyReLU, Alpha:0.2 \\ 
 & Dropout & (128,) & 0 & Regularization, Rate:0.15 \\
 & Dense & (10,) & 92170 & Filter:10 \\
 & Softmax & (10,) & 0 & Activation: softmax \\ 
 \hline
\end{tabular}
\end{table}
}


\newpage

\bibliography{ref}{}
\bibliographystyle{aasjournal}



\end{document}